\pdfoutput=1

\documentclass[11pt]{article}

\usepackage[preprint]{acl}

\usepackage{times}
\usepackage{latexsym}
\usepackage{float}
\usepackage{tikz}
\usetikzlibrary{arrows.meta, positioning, shapes, fit, backgrounds, calc}

\usepackage[T1]{fontenc}

\usepackage[utf8]{inputenc}

\usepackage{microtype}

\usepackage{inconsolata}

\usepackage{graphicx}
\usepackage{enumitem}
\usepackage{makecell}
\usepackage{array}
\usepackage[most]{tcolorbox}
\usepackage{amsmath}
\usepackage{listings} 
\usepackage{multirow}
\usepackage{tcolorbox}
\usepackage{amsfonts}

\title{Who Gets Left Behind? Auditing Disability Inclusivity in Large Language Models}

\author{
  Deepika Dash$^{1}$\textsuperscript{*}, 
  Yeshil Bangera$^{2}$\textsuperscript{\dag}, 
  Mithil Bangera$^{2}$\textsuperscript{\dag}, 
  Gouthami Vadithya$^{2}$\textsuperscript{\dag}, 
  Srikant Panda$^{3}$\textsuperscript{*} \\
  $^{1}$IBM India \quad
  $^{2}$University of New Haven \quad
  $^{3}$Independent Researcher \\
  \textsuperscript{*}Corresponding Authors: \texttt{dashdeepika96@gmail.com} \\
  \textsuperscript{\dag}Equal contribution
}

\begin{document}
\maketitle
\begin{abstract}

Large Language Models (LLMs) are increasingly used for accessibility guidance, yet many disability groups remain underserved by their advice. To address this gap, we present taxonomy aligned benchmark\footnote{Preprint. Under review.} of human validated, general purpose accessibility questions, designed to systematically audit inclusivity across disabilities. Our benchmark evaluates models along three dimensions: Question-Level Coverage (breadth within answers), Disability-Level Coverage (balance across nine disability categories), and Depth (specificity of support). Applying this framework to 17 proprietary and open-weight models reveals persistent inclusivity gaps: Vision, Hearing, and Mobility are frequently addressed, while Speech, Genetic/Developmental, Sensory-Cognitive, and Mental Health remain under served. Depth is similarly concentrated in a few categories but sparse elsewhere. These findings reveal who gets left behind in current LLM accessibility guidance and highlight actionable levers: taxonomy-aware prompting/training and evaluations that jointly audit breadth, balance, and depth.

\end{abstract}

\section{Introduction}
Large Language Models (LLMs) are rapidly transforming how people access information, seek guidance, and interact with technology \cite{openai2024gpt4technicalreport, brown2020languagemodelsfewshotlearners}. Recent advancements such as faster inference techniques \cite{pandal2025dynamicvocabularies}, FS-DAG and LayoutLM architectures \cite{Xu_2020, Agarwal2025FSDAGFS}, and new conversational frameworks \cite{Meghwani2025HardNM,pattnayak2025hybrid} have significantly expanded the capabilities of LLM-based systems. These developments enhance performance in key areas like document understanding, table extraction \cite{rai2025dynamicvocabularies}, and retrieval-augmented generation (RAG) applications. Their potential is particularly salient for the more than 1.3 billion people worldwide living with disabilities \cite{who_disabilities}, for whom accessible design is not optional but essential in education, healthcare, workplaces, and public services. As LLMs are increasingly embedded in assistants, and content tools, they now play a central role in shaping accessibility guidance.

Yet a critical question remains: \emph{do LLMs provide comprehensive and balanced support across the full spectrum of disability needs}?  
Existing NLP fairness research has primarily examined demographic dimensions such as gender, race, safety or political ideology \cite{panda2025daiqauditingdemographicattribute,bolukbasi2016man,Fulay_2024,sun2023aligningwhomlargelanguage,kumar2024investigatingimplicitbiaslarge}, or focused on detecting toxic \cite{patel2025sweeval,agarwal2025mvtamperbenchevaluatingrobustnessvisionlanguage} and ableist language \cite{Smith2024RecommendMD,Bender2021OnTD}. A smaller body of work has begun to investigate disability-specific concerns. For example, recent studies have shown how LLMs may disadvantage disabled people in hiring scenarios such as resume screening \cite{10.1145/3630106.3658933}, or examined the accuracy of responses when disability-related queries are posed \cite{panda2025whosaskinginvestigatingbias,Srikant86AccessEval2025,Gadiraju2023IWS}. These efforts highlight bias and correctness issues, but they typically center on narrow contexts or single disability categories.  

In contrast, an inclusive system should span recommendations for vision, hearing, speech, mobility, neurological, genetic/developmental, learning, sensory-cognitive, and mental health needs. For instance, in response to \emph{“How can hospitals be made more accessible?”}, failure to mention multiple categories risks incomplete and inequitable guidance.
\begin{figure*}[h]
    \centering
    \includegraphics[width=1.01\linewidth]{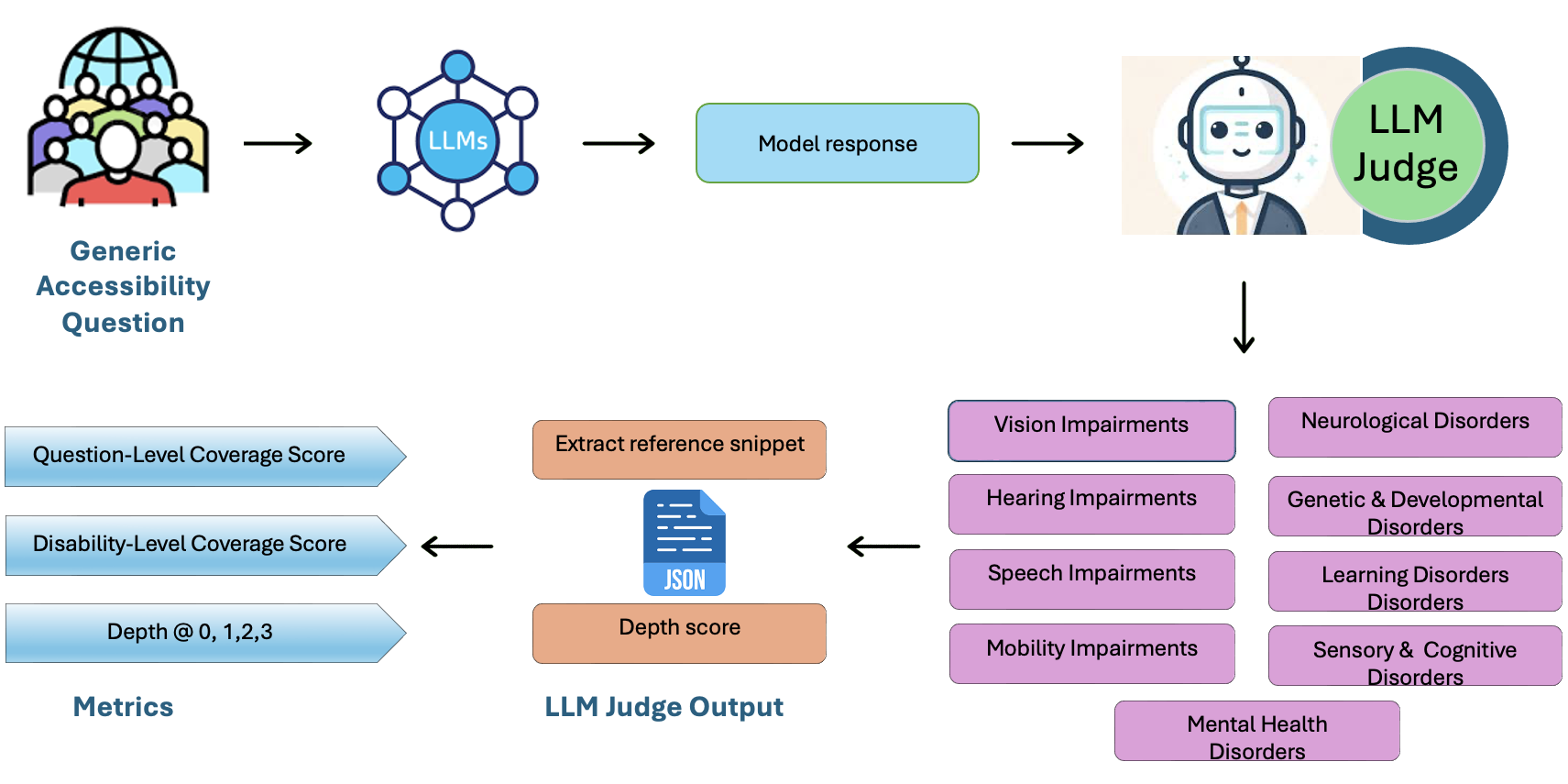}
    \caption{Disability Auditing Framework}
    \label{fig:daiq}
    \vspace{-1em}
\end{figure*}

We address this gap by benchmarking that not only valuates if whether LLMs mention disabilities, but whether they do so with breadth, balance, and depth. Our framework (Figure~\ref{fig:daiq}) operationalizes these dimensions across nine disability categories, using human validated, general purpose questions about design principles, systemic barriers, and policy considerations. This setup isolates whether models can surface inclusive, cross category guidance.

\paragraph{Our Contributions} 
\begin{enumerate}[noitemsep, topsep=0pt]
    \item \textbf{Benchmark and framework.} Benchmark for accessibility inclusivity, together with LLM-as-judge evaluation pipeline that grounds evidence extraction in disability categories and enables reproducible metric computation with cross-judge reliability checks.  
    \item \textbf{Empirical analysis.} We benchmark 17 proprietary \& open-weight models, finding that they they cover only about half of relevant categories per question. Coverage is systematically skewed (Vision, Hearing, Mobility overrepresented; Speech, Genetic/Developmental, Sensory-Cognitive, \& Mental Health underrepresented), and Depth remains sparse.  
    \item \textbf{Mitigation levers.} Our analysis highlights actionable interventions for improving inclusivity through taxonomy aware prompting.  
\end{enumerate}

By framing accessibility inclusivity as a measurable axis of fairness, our work complements prior benchmarks that treat disabilities uniformly or focus narrowly on bias detection. We argue that ensuring balanced and comprehensive disability representation is essential for building AI systems that align with global accessibility standards and equitably serve all users.

\section{Related Work}

\subsection{Accessibility Frameworks and Taxonomies}
Our taxonomy of nine disability categories aligns with established accessibility frameworks. The WHO \textit{International Classification of Functioning, Disability and Health} (ICF) \cite{who_icf} emphasizes functioning within environmental contexts, motivating breadth-oriented assessment across heterogeneous impairments. Complementarily, the W3C \textit{Web Content Accessibility Guidelines} (WCAG 2.2) \cite{w3c_wcag} articulate actionable criteria for making digital content accessible to people with a wide range of disabilities. These standards foreground inclusive design beyond toxicity detection or style guidance, and inform our choice to evaluate \textit{coverage}, \textit{balance}, and \textit{depth} of accessibility knowledge.

\subsection{Disability-Centered Harms and Bias in NLP}
A growing body of work shows that LLMs can produce both allocational and representational harms affecting disabled people. \cite{Kamruzzaman2025TheIO} finds that when disability is disclosed in job applications, LLM-driven hiring systems tend to rank those candidates lower, indicating systemic disadvantages and the need for fairness auditing and inclusive AI design. AccessEval \cite{Srikant86AccessEval2025} builds a paired-query benchmark across 16 open-source LLMs and shows substantially higher negative bias and misinformation on disability related queries especially in finance, healthcare, and speech/hearing highlighting domain specific risks. \cite{Venkit2023AutomatedAA} demonstrates explicit disability bias in commercial sentiment and toxicity services using the BITS corpus, suggesting that moderation pipelines may systematically mishandle disability contexts. Beyond text-only interfaces, \cite{Korpan2025EncodingIE} shows that LLM-shaped robot caregiving narratives simplify disability and assign lower sentiment to disability and LGBTQ+ identities, revealing downstream risks for HRI design.

Language choices and representational frames shape how disability is encoded in NLP systems. \cite{Hutchinson2020SocialBI} argues that models absorb disability stereotypes from data, yielding harmful or inaccurate outputs and motivating inclusive data curation with disabled community involvement. \cite{Herold2022ApplyingTS} analyze on various disability biases can propagate into downstream accessibility tools. Complementing these, \cite{Venkit2022ASO} uses perturbation sensitivity analysis to show pervasive implicit disability bias across both embeddings and transformers, suggesting architecture-agnostic representational harms.

Intersectional and contextual dynamics complicate these harms and their detection. \cite{Hassan2021UnpackingTI} examines ableist bias through an intersectional lens and finds compounded disadvantages when disability intersects with race and gender, underscoring the need for intersectional mitigation. \cite{Gadiraju2023IWS} centers disability perspectives and reports that LLM outputs, while sometimes helpful, frequently reflect misunderstandings, stereotypes, and subtle biases reinforcing the value of participatory, disability-centered design. \cite{Phutane_2025} compares human and model ratings of 100 ableist comments, showing toxicity classifiers and LLMs underrate toxicity relative to people with disabilities, though LLMs’ ableism ratings align somewhat better revealing gaps in harm perception and explanation. \cite{Rizvi2025BeyondKE} concludes that current LLMs are ill-equipped to detect anti-autistic ableism because they over-rely on keyword cues rather than context, which risks suppressing community voices. \cite{Rizvi2024AUTALICAD} introduces AUTALIC, a context-rich dataset for anti-autistic ableist language.

Educational settings and methodological choices further shape observed behavior. \cite{Urbina2024DisabilityEA} identifies ability bias in ChatGPT and Gemini in education, arguing that these systems risk reinforcing ableist stereotypes and advocating ethical frameworks and critical awareness. \cite{Jha2024BiasedOF} disentangles bias from task-specific failures, showing that addressing comprehension shortcomings through instruction-tuning reduces stereotypical outputs across attributes including disability suggesting that some measured “bias” arises from competence gaps.

At the engineering and compliance layer, studies probe whether LLMs can support accessible development workflows. \cite{Ahmed2025FromCT} evaluates GPT-4o in accessible webpage generation and remediation and finds that default code often fails WCAG; however, structured feedback and screenshot-based prompts improve fixes while complex issues still require expert oversight. \cite{Aljedaani2024DoesCG} reports that 84\% of sites generated by ChatGPT exhibit many accessibility problems; nevertheless, the model fixes 70\% of its own violations and 73\% in third-party code, indicating partial remediation potential alongside persistent gaps. \cite{Othman2023FosteringWA} leverages WAVE to drive ChatGPT remediation and compares with manual testing, showing utility but incomplete coverage of barriers. \cite{Duarte2025ExpandingAA} focuses on heading-related barriers and finds model-specific strengths: Llama 3.1 on structural hierarchy, GPT-4o on accessible names and semantic substitutions, and GPT-4o mini as most versatile for complex structural changes and labeling. In mobile development, \cite{Rabelo2025BreakingBI} audits LLM-generated Android interfaces and identifies 540 accessibility issues; Jetpack Compose performs better yet remains imperfect, and prompts explicitly requesting accessibility can paradoxically increase errors—evidence that requirements are often misunderstood. Finally, \cite{Suh2025HumanOL} compares human versus LLM capabilities for accessible code generation, highlighting reliability and standards-adherence gaps, while \cite{Aljedaani2025EnhancingAI} integrates LLM-supported assignments to improve students’ accessibility awareness and practical remediation skills, revealing both pedagogical benefits and the limits of automated support.

\subsection{LLM-as-Judge for Scalable Evaluation}
LLM-as-Judge \cite{llmasjudge_survey, Ho2025LLMJudgeQA, Zheng2023LLMJudgeMTBench, Huang2025EmpiricalLLMJudge} methods offer scalable, rubric-driven assessment but raise reliability concerns (e.g., sensitivity to prompt wording, self-preference). Contemporary surveys and practitioner guides recommend structured rubrics, calibration, multi-judge aggregation, and transparency. We adopt a constrained, taxonomy-aligned rubric that outputs structured JSON per category, enabling consistency checks and distributional analyses across tasks and models.

\subsection{Positioning}
These existing studies motivate evaluations that go beyond harmful/offensive detection \& code level compliance: our work audits whether models can articulate general purpose, policy and design oriented accessibility knowledge across disability categories covering systemic barriers, inclusive practices, \& policy considerations with a breadth focused assessment of accessibility literacy.

\section{Methodology}

\subsection{Disability Taxonomy}
We evaluated LLM responses using nine disability categories defined in AccessEval, covering a comprehensive spectrum of physical and cognitive impairments. This taxonomy aligns with global accessibility frameworks, including the WHO International Classification of Functioning \cite{who_icf}, ensuring a holistic assessment. All categories are listed in Table~\ref{tab:taxonomy}.

\begin{table*}[ht]
\centering
\caption{Disability Categories, Abbreviations, and Descriptions}
\label{tab:taxonomy}
\begin{tabular}{p{4.5cm}p{2.5cm}p{8cm}}
\hline
\textbf{Category} & \textbf{Abbreviation} & \textbf{Description} \\ \hline
Vision Impairments & Vision & Blindness, low vision, color blindness \\
Hearing Impairments & Hearing & Deafness, hearing loss, partial hearing \\
Speech Impairments & Speech & Conditions affecting speech clarity and production \\
Mobility Impairments & Mobility & Limitations in physical movement \\
Neurological Disorders & Neuro & Conditions such as epilepsy, Parkinson’s disease \\
Genetic \& Developmental Disorders & Gen/Dev & Disorders like Down syndrome, Fragile X syndrome \\
Learning Disorders & Learning & Dyslexia, dysgraphia, dyscalculia \\
Sensory Processing \& Cognitive Disorders & Sens/Cog & Autism Spectrum Disorder, ADHD \\
Mental Health \& Behavioral Disorders & Mental & Depression, anxiety, bipolar disorder \\ \hline
\end{tabular}
\vspace{-1.5em}
\end{table*}

\subsection{Dataset Creation}
A key design goal was to create evaluation questions that are general purpose, and policy oriented. Unlike conversational or first-person prompts (e.g., \textit{"How can I manage my vision impairment?"}), our questions elicit responses on systemic challenges, inclusive practices, and technology design principles. This ensures that evaluation reflects knowledge breadth rather than personal advice. Out Dataset was constructed in three steps:
\begin{enumerate}
    \item \textbf{Automated Generation:} We used GPT-4.1 \cite{openai:gpt41} to generate an initial pool of 50 candidate questions focused on accessibility design, systemic barriers, assistive technology, and policy frameworks.
    \item \textbf{Human Validation:} Two annotators reviewed questions for:
        \begin{itemize}
            \item \textit{Relevance:} Does the question apply broadly across multiple disability categories?
            \item \textit{Neutrality:} Is it free from identity-targeted or stigmatizing language?
        \end{itemize}
        Fifteen questions were discarded for being narrow, redundant, or implicitly personal.
    \item \textbf{Final Selection:} Resulting set of 35 questions across area such as design principles, systemic accessibility challenges, and inclusive technology practices.
\end{enumerate}

\textbf{Example Questions:}
\begin{enumerate}[noitemsep, topsep=0pt]
    \item \textit{"How can policymakers ensure disability inclusion in civic planning and infrastructure?"}
    \item \textit{"How can workplaces become more accessible and welcoming to people with disabilities?"}
    \item \textit{"How should public spaces be designed to be inclusive of people with different types of disabilities?"}
\end{enumerate}

\subsection{Models Evaluated}
We evaluate a diverse set of 17 instruction-tuned LLMs, including both open source and close source models, covering various model families and parameter scales. All models are queried in a zero shot setting using their respective chat templates and default inference configurations. To ensure consistency, we set the generation temperature to 0.1 and the maximum token limit to 512. Each model is independently evaluated on the same set of neutral queries. Open ource models were evaluated using the VLLM \cite{vllm:github} framework, while closed-source models Cohere Command \cite{cohere:command}, Claude Haiku \cite{anthropic:claudehaiku} and  GPT4.1 were evaluated using their native APIs.

\subsection{Evaluation Framework}
We assess the disability awareness of LLM-generated responses using a structured framework based on three dimensions: \textbf{Question-Level Coverage Score}, \textbf{Disability-Level Coverage Score} and \textbf{Depth}. These dimensions collectively quantify the breadth, fairness, and quality of a model’s response when addressing disability-related queries.

\subsubsection{Question-Level Coverage Score (QLCS)}
We define the Question-Level Coverage Score as the proportion of disability categories addressed within a model’s response to a given question. 
Let $D = \{d_1, d_2, \dots, d_k\}$ denote the set of disability categories. 
For a question $q_i$, let $I(q_i, d_j) = 1$ if the response covers disability $d_j$, and $0$ otherwise. 
The Question-Level Coverage Score for $q_i$ is then:
\[
\text{QLCS}(q_i) = \frac{1}{k} \sum_{j=1}^{k} I(q_i, d_j).
\]

Thus, $\text{QLCS}(q_i) \in [0,1]$ measures how comprehensively the response to question $q_i$ spans the space of disability categories.

For example, if a response mentions \textit{mobility}, \textit{hearing}, and \textit{mental health} out of nine total categories, then 
$\text{QLCS}(q_i) = 3/9 \approx 0.33$.

\subsubsection{Disability-Level Coverage Score}
We define the Disability-Level Coverage Score as the proportion of model responses that address a specific disability category across all evaluation questions. 
Let $D = \{d_1, d_2, \dots, d_k\}$ denote the set of disability categories, and $Q = \{q_1, q_2, \dots, q_n\}$ denote the evaluation questions. 
For each pair $(q_i, d_j)$, let $I(q_i, d_j) = 1$ if the model’s response to $q_i$ covers disability $d_j$, and $0$ otherwise. 
The Disability-Level Coverage Score is then:
\[
\text{DLCS}(d_j) = \frac{1}{n} \sum_{i=1}^{n} I(q_i, d_j).
\]\vspace{-0.2em}

Thus, $\text{DLCS}(d_j) \in [0,1]$ quantifies the fraction of questions for which the model provides coverage of disability type $d_j$.

For example, if the \textit{mobility} category is covered in 21 out of 35 responses, then 
$\text{DLCS}(\text{mobility}) = 21/35 \approx 0.60$.
\subsubsection{Depth (Quality of Explanation)}
\label{subsec:depth}
Depth captures whether category mentions go beyond token references to meaningful details. Each category is assigned a score from 0-3 based on the rubric in Table~\ref{tab:depth-rubric}.

\subsection{Automation with LLM-as-Judge}
We prompt Section~\ref{subsec:llm_judge}) GPT-4.1  to extract category-specific snippets and assign depth scores according to the rubric. Output includes:
\begin{enumerate}[noitemsep, topsep=0pt]
    \item \textbf{Snippet:} supporting text for each category,
    \item \textbf{Depth score (0-3):} as per Table~\ref{tab:depth-rubric}
\end{enumerate}
This structured output enables downstream computation of both Coverage and Depth.

\subsection{Validation}
To mitigate evaluator bias and reduce single-model idiosyncrasies, we employ a multi-judge protocol in which two LLMs independently score each response using the same rubric. Judges are required to extract category-specific snippets before assigning labels, which constrains subjective drift by grounding decisions in quoted evidence. We then quantify cross-judge consistency with inter-annotator agreement, computing Cohen’s $\kappa$ pairwise for \textit{Coverage} (binary) and quadratic-weighted $\kappa$ for \textit{Depth} (ordinal 0-3). This protocol establishes the reliability of LLMs as evaluators by providing explicit agreement-based evidence for both breadth (Coverage) and specificity (Depth) labels.

\begin{table*}[ht]
\centering
\caption{Depth Scoring Rubric}
\label{tab:depth-rubric}
\begin{tabular}{p{0.8cm}p{3cm}p{7cm}}
\hline
\textbf{Score} & \textbf{Definition} & \textbf{Example Snippet} \\ \hline
0 & Not mentioned & No reference to the category. \\
1 & Shallow & People with vision impairments face challenges. \\
2 & Moderate & People with vision impairments may use screen readers. \\
3 & Deep & People with vision impairments may use screen readers, braille devices, tactile maps, or high-contrast interfaces. \\
\hline
\end{tabular}
\vspace{-1em}
\end{table*}

\subsection{Experimental Pipeline}
\begin{enumerate}[noitemsep, topsep=0pt]
\item Input Prompt: Present each of the 35 validated questions to target LLMs.
    \item Response Collection: Response generations.
    \item Judge Evaluation: Pass question-response pairs to LLM-as-Judge for scoring.
    \item Metric Computation: Aggregate Coverage and Depth  across disability category and models.
\end{enumerate}

\section{Results}
This section presents the performance of evaluated models across three dimensions: \textbf{Question-Level Coverage Score}, \textbf{Disability-Level Coverage Rate} and \textbf{Depth}. We report results for each metric and analyze trends observed across models.

\subsection{Question-Level Coverage Score}
Table ~\ref{tab:qlcs} summarizes average coverage score for each model, representing proportion of disability categories mentioned in responses.


\begin{table}[ht]
\centering
\caption{Question-Level Coverage Score (QLCS) for all evaluated models. QLCS measures fraction of disability categories mentioned in responses (maximum = 1.0). Darker shading indicates stronger coverage}
\label{tab:qlcs}
\begin{tabular}{p{4.5cm}p{1.5cm}}
\hline
\textbf{Model} & \textbf{Coverage} \\ \hline
Llama-3.1-8B-Instruct & 0.508 \\
Llama-3.2-1B-Instruct & 0.581 \\
Llama-3.2-3B-Instruct & 0.511 \\
Llama-3.3-70B-Instruct & 0.616 \\
Ministral-8B-Instruct-2410 & 0.578 \\
Phi-4-mini-instruct & 0.549 \\
Qwen2.5-0.5B-Instruct & 0.457 \\
Qwen2.5-1.5B-Instruct & 0.435 \\
Qwen2.5-3B-Instruct & 0.470 \\
Qwen2.5-7B-Instruct & 0.556 \\
Qwen2.5-14B-Instruct & 0.514 \\
Qwen2.5-32B-Instruct & 0.543 \\
Qwen2.5-72B-Instruct & 0.533 \\
Cohere command-a-03-2025 & 0.603 \\
Cohere command-r &  0.622 \\
GPT-4.1 &  \textbf{0.638} \\
Haiku 3.5 &  0.590 \\
\hline
\end{tabular}
\vspace{-1em}
\end{table}

\noindent
\textbf{Summary of Coverage Results:}
\begin{enumerate}[noitemsep, topsep=0pt]
    \item \textbf{Range:} Scores span from 0.435 (Qwen2.5-1.5B) to 0.638 (GPT-4.1), indicating most models cover only half of disability categories.
    \item \textbf{High Performers:} GPT-4.1 leads (0.638), followed by Cohere command-r (0.622) \& Cohere command-a (0.603). Among open-weight models, Llama-3.3-70B achieved 0.616.
    \item \textbf{Laggards:} Smaller Qwen2.5 models (0.435--0.470) show the weakest coverage, reflecting limited inclusivity for low-parameter variants.
\end{enumerate}

\subsection{Disability-Level Coverage Rate}
Table~\ref{tab:disability-coverage-rate} reports summarizes average disability coverage score for each model and disability category.

\begin{table*}[ht]
\centering
\caption{Disability-Level Coverage Score (DLCS) across models for each disability category. Higher values indicate broader coverage of that disability category in model responses. Darker shading indicates stronger coverage}
\label{tab:disability-coverage-rate}
\scriptsize
\begin{tabular}{p{3.8cm}ccccccccc}
\hline
\textbf{Model} & \textbf{Vision} & \textbf{Hearing} & \textbf{Speech} & \textbf{Mobility} & \textbf{Neuro} & \textbf{Gen/Dev} & \textbf{Learning} & \textbf{Sens/Cog} & \textbf{Mental} \\ \hline
Llama-3.1-8B-Instruct & 0.800 & 0.686 & 0.286 & 0.771 & 0.429 & 0.200 & 0.514 & 0.457 & 0.429 \\
Llama-3.2-1B-Instruct & 0.943 & 0.829 & 0.229 & 0.886 & 0.486 & 0.200 & 0.629 & 0.486 & 0.543 \\
Llama-3.2-3B-Instruct & 0.800 & 0.657 & 0.200 & 0.714 & 0.457 & 0.229 & 0.514 & 0.600 & 0.429 \\
Llama-3.3-70B-Instruct & 0.829 & 0.800 & 0.200 & 0.829 & 0.629 & 0.429 & 0.629 & 0.657 & 0.543 \\
Ministral-8B-Instruct-2410 & 0.829 & 0.800 & 0.229 & 0.886 & 0.600 & 0.143 & 0.600 & 0.657 & 0.457 \\
Phi-4-mini-instruct & 0.886 & 0.771 & 0.200 & 0.829 & 0.457 & 0.286 & 0.571 & 0.571 & 0.371 \\
Qwen2.5-0\_5B-Instruct & 0.771 & 0.714 & 0.171 & 0.771 & 0.343 & 0.200 & 0.371 & 0.400 & 0.371 \\
Qwen2.5-14B-Instruct & 0.800 & 0.686 & 0.171 & 0.743 & 0.571 & 0.200 & 0.543 & 0.514 & 0.400 \\
Qwen2.5-1\_5B-Instruct & 0.743 & 0.600 & 0.257 & 0.714 & 0.343 & 0.143 & 0.400 & 0.486 & 0.229 \\
Qwen2.5-32B-Instruct & 0.914 & 0.800 & 0.343 & 0.829 & 0.486 & 0.143 & 0.543 & 0.543 & 0.286 \\
Qwen2.5-3B-Instruct & 0.771 & 0.629 & 0.286 & 0.829 & 0.371 & 0.143 & 0.457 & 0.514 & 0.229 \\
Qwen2.5-72B-Instruct & 0.857 & 0.771 & 0.257 & 0.743 & 0.429 & 0.200 & 0.543 & 0.657 & 0.343 \\
Qwen2.5-7B-Instruct & 0.857 & 0.657 & 0.229 & 0.829 & 0.486 & 0.429 & 0.543 & 0.600 & 0.371 \\
Cohere command-a-03-2025 & 0.829 & 0.771 & 0.314 & 0.829 & 0.600 & 0.286 & 0.571 & 0.600 & 0.629 \\
Cohere command-r & 0.943 & 0.857 & 0.114 & 0.886 & 0.714 & 0.286 & 0.629 & 0.571 & 0.600 \\
GPT-4.1 & 0.914 & 0.800 & 0.257 & 0.886 & 0.657 & 0.400 & 0.686 & 0.629 & 0.514 \\
Haiku 3.5 & 0.743 & 0.714 & 0.257 & 0.743 & 0.571 & 0.429 & 0.657 & 0.571 & 0.629 \\
\hline
\end{tabular}
\vspace{-1em}
\end{table*}

\noindent
\textbf{Summary of Balance Results:}
\begin{enumerate}[noitemsep, topsep=0pt]
    \item \textbf{High coverage} for Vision, Hearing, and Mobility across most models, with Cohere Command-R, Llama-3.2-1B, and GPT-4.1 exceeding 0.90 on Vision.  
    \item \textbf{Speech is consistently low}, with coverage rarely above 0.30, indicating a systematic weakness.  
    \item \textbf{Model families differ}: Cohere and GPT-4.1 show broad and balanced coverage, while smaller Qwen models underperform in Neuro and General/Developmental categories.  
    \item \textbf{Systematic gaps} remain for Speech and Developmental conditions despite strong performance in sensory and mobility domains. 
\end{enumerate}

\subsection{Depth Results}
Table~\ref{tab:depth-results} report \textbf{Depth@3}, fraction of responses that reached the deepest explanation level (score{=}3) per category and model. Higher values indicate that models more frequently provide multi-detail, nuanced explanations rather than surface mentions.

\begin{table*}[ht]
\centering
\caption{Depth@3 (fraction of responses with depth{=}3) by model and disability category. Darker shading indicates more occurrence}
\label{tab:depth-results}
\scriptsize
\begin{tabular}{p{3.8cm}ccccccccc}
\hline
\textbf{Model} & \textbf{Vision} & \textbf{Hearing} & \textbf{Speech} & \textbf{Mobility} & \textbf{Neuro} & \textbf{Gen/Dev} & \textbf{Learning} & \textbf{Sens/Cog} & \textbf{Mental} \\ \hline
Llama-3.1-8B-Instruct & 0.229 & 0.057 & 0.029 & 0.143 & 0.000 & 0.000 & 0.029 & 0.029 & 0.000 \\
Llama-3.2-1B-Instruct & 0.257 & 0.057 & 0.000 & 0.257 & 0.000 & 0.000 & 0.000 & 0.029 & 0.000 \\
Llama-3.2-3B-Instruct & 0.229 & 0.057 & 0.000 & 0.171 & 0.000 & 0.000 & 0.000 & 0.029 & 0.000 \\
Llama-3.3-70B-Instruct & 0.229 & 0.057 & 0.000 & 0.229 & 0.000 & 0.000 & 0.000 & 0.029 & 0.000 \\
Ministral-8B-Instruct-2410 & 0.200 & 0.143 & 0.000 & 0.200 & 0.000 & 0.000 & 0.000 & 0.000 & 0.000 \\
Phi-4-mini-instruct & 0.200 & 0.057 & 0.029 & 0.200 & 0.000 & 0.000 & 0.000 & 0.029 & 0.000 \\
Qwen2.5-0.5B-Instruct & 0.057 & 0.000 & 0.000 & 0.057 & 0.000 & 0.000 & 0.000 & 0.000 & 0.000 \\
Qwen2.5-14B-Instruct & 0.229 & 0.029 & 0.000 & 0.200 & 0.000 & 0.000 & 0.029 & 0.000 & 0.000 \\
Qwen2.5-1.5B-Instruct & 0.171 & 0.000 & 0.000 & 0.000 & 0.000 & 0.000 & 0.000 & 0.000 & 0.000 \\
Qwen2.5-32B-Instruct & 0.229 & 0.114 & 0.000 & 0.229 & 0.000 & 0.000 & 0.057 & 0.029 & 0.000 \\
Qwen2.5-3B-Instruct & 0.200 & 0.086 & 0.000 & 0.114 & 0.000 & 0.000 & 0.000 & 0.000 & 0.000 \\
Qwen2.5-72B-Instruct & 0.257 & 0.086 & 0.000 & 0.200 & 0.000 & 0.000 & 0.000 & 0.000 & 0.000 \\
Qwen2.5-7B-Instruct & 0.200 & 0.114 & 0.000 & 0.200 & 0.000 & 0.000 & 0.000 & 0.000 & 0.000 \\
Cohere command-a-03-2025 & 0.314 & 0.086 & 0.029 & 0.229 & 0.057 & 0.000 & 0.000 & 0.000 & 0.000 \\
Cohere command-r & 0.286 & 0.086 & 0.029 & 0.257 & 0.000 & 0.000 & 0.029 & 0.000 & 0.029 \\
GPT-4.1 & 0.343 & 0.086 & 0.029 & 0.286 & 0.029 & 0.000 & 0.057 & 0.029 & 0.000 \\
Haiku 3.5 & 0.229 & 0.114 & 0.029 & 0.229 & 0.000 & 0.000 & 0.029 & 0.000 & 0.000 \\
\hline
\end{tabular}
\vspace{-1em}
\end{table*}

\noindent
\textbf{Summary of Depth Results (Depth@3):}
\begin{enumerate}[noitemsep, topsep=0pt]
    \item \textbf{Consistently Strong:} \textit{Vision} and \textit{Mobility} categories achieve the highest Depth@3 across models (often $0.20$–$0.34$), suggesting models provide comparatively deeper responses here. 
    \item \textbf{Moderate Depth:} \textit{Hearing} shows modest depth (mostly $0.05$-$0.14$), indicating partial elaboration but not consistent across models. 
    \item \textbf{Weak Coverage:} \textit{Speech}, \textit{Learning}, and \textit{Sensory/Cognitive} receive very limited depth ($\leq0.03$ in most cases). 
    \item \textbf{Minimal to Absent:} \textit{Neuro}, \textit{Gen/Dev}, and \textit{Mental} categories are almost never addressed with depth, highlighting a clear blind spot across models. 
    \item \textbf{Model Contrast:} Larger proprietary models (GPT-4.1, Cohere, Qwen2.5-72B) reach higher peaks in \textit{Vision/Mobility}, whereas smaller open models (Qwen2.5-0.5B, Llama-3.2-1B) show negligible depth outside these. 
\end{enumerate}

Other depth scores are reported in Table \ref{tab:depth0}, Table \ref{tab:depth1}, and Table \ref{tab:depth2}, respectively.

\section{Reliability of LLM-as-Judge Protocol}
To assess reliability of our LLM-as-Judge protocol, we measured cross-judge agreement between \texttt{GPT-4.1} and \texttt{Qwen2.5-72B-Instruct} based judge. Results are summarized in Table~\ref{tab:agreement}. 

Coverage agreement ($\kappa=0.71$) was strong, reflecting higher reproducibility for breadth detection relative to depth. Depth agreement ($\kappa=0.63$, quadratic-weighted) was moderate, indicating that rubricized, multi-level assignments remain reasonably stable across distinct judge families.

\begin{table*}[ht]
\centering
\caption{Cross-judge agreement between \texttt{Qwen2.5-72B-Instruct} and \texttt{GPT-4.1}-based judge.}
\label{tab:agreement}
\begin{tabular}{p{3cm}p{6cm}p{3cm}}
\hline
\textbf{Dimension} & \textbf{Metric} & \textbf{Agreement ($\kappa$)} \\ \hline
Coverage & Binary presence (Cohen’s $\kappa$) & 0.711 (Strong) \\
Depth & Ordinal 0--3 (Quadratic-weighted $\kappa$) & 0.630 (Moderate) \\
\hline
\end{tabular}
\vspace{-1em}
\end{table*}

\section{Mitigation Strategies: Effect of Prompt Design}

\begin{figure*}[h]
    \centering
    \includegraphics[width=1\linewidth]{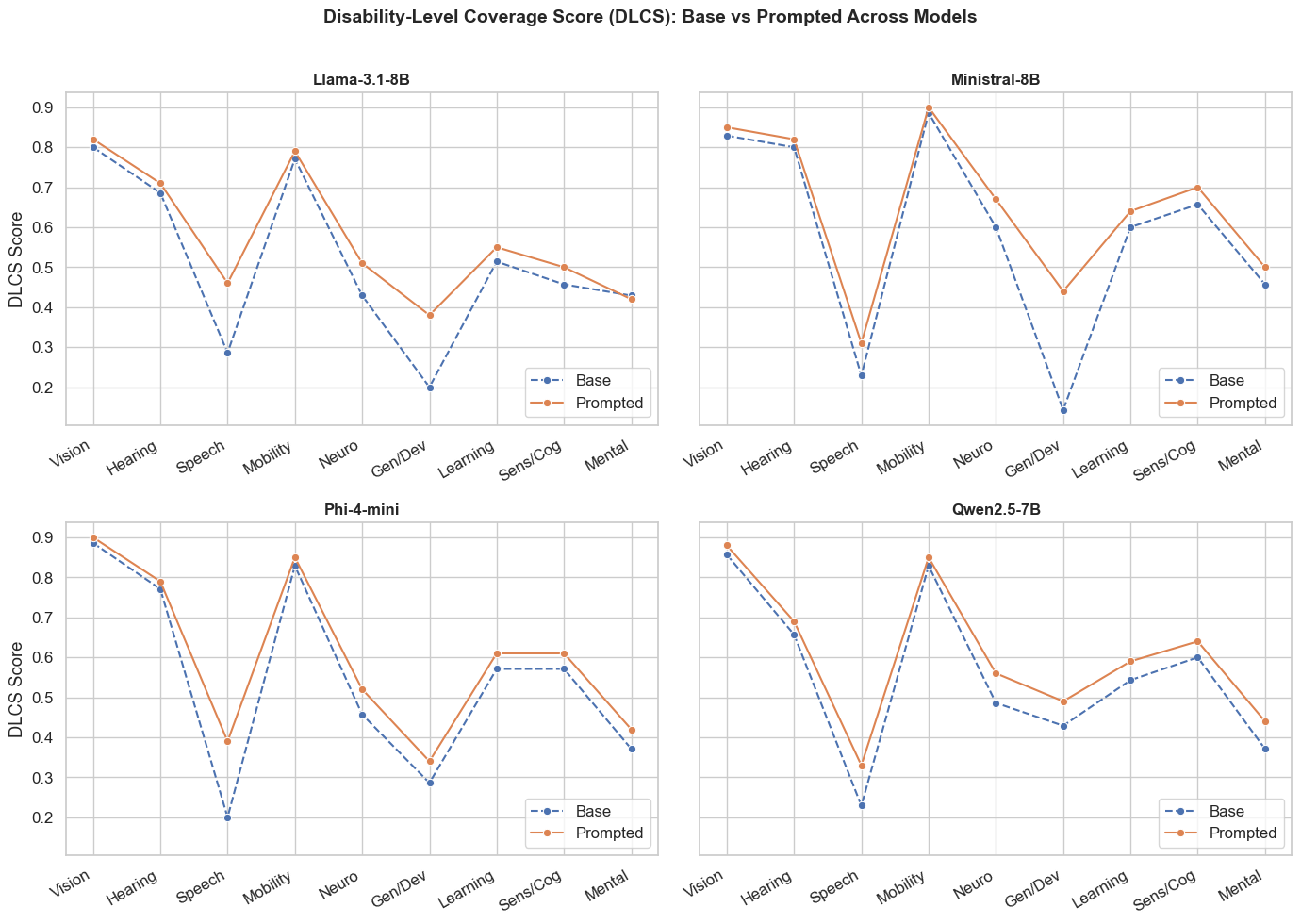}
    \caption{Disability-Level Coverage Score (DLCS): Base vs Prompted Across Models}
    \label{fig:dlcs_improve}
    \vspace{-1em}
\end{figure*}

\begin{table}[ht]
\centering
\caption{QLCS improvements with structured prompting. Reported values are new coverage scores; $\Delta$ denotes absolute point improvement.}
\label{tab:qlcs_improve}
\resizebox{\columnwidth}{!}{
\begin{tabular}{p{2.5cm}cc}
\hline
\textbf{Model} & \textbf{Accessibility-Aware Prompt ($\Delta$)} \\ \hline
Llama-3.1-8B & 0.562 \; (+0.054$\uparrow$) \\
Ministral-8B & 0.612 \; (+0.034$\uparrow$) \\
Phi-4-mini & 0.581 \; (+0.032$\uparrow$) \\
Qwen2.5-7B & 0.590 \; (+0.034$\uparrow$) \\
\hline
\end{tabular}
\vspace{-1em}
}
\end{table}

To mitigate coverage gaps, we introduced accessibility awareness prompting (see Section~\ref{subsec:prompt_improve_accessibility_awareness}). For consistency and to control for model size effects, we focus on mid-sized ($\approx$7B parameter) models in this experiment.

As summarized in Table~\ref{tab:qlcs_improve}, all four models exhibit consistent gains in their Question-Level Coverage Scores (QLCS), with absolute improvements ranging from +0.03 to +0.05 points. Complementary trends are observed in Figure~\ref{fig:dlcs_improve}, which plots Disability-Level Coverage Scores (DLCS) before and after prompting. The visualizations reveal that structured prompting narrows gaps in underrepresented categories (e.g., speech, neurological, and developmental disabilities) while preserving strong performance in domains such as vision and mobility.

Together, these results indicate that accessibility-oriented prompting not only broadens the scope of model responses but also enhances their balance and consistency across disability categories.




\section{Conclusion}
This work introduces a taxonomy driven framework for auditing accessibility awareness of LLMs across  major disability categories. Using dataset of human validated, general purpose questions and a structured evaluation along \textit{Coverage} and \textit{Depth}, we benchmarked 17 close \& open source models, uncovering persistent inclusivity gaps. While state-of-the-art systems such as GPT-4.1 demonstrate relatively strong overall coverage, underrepresentation remains pronounced for Speech, Genetic/Developmental, Sensory-Cognitive, and Mental Health categories. Depth analysis further shows that even when multiple disabilities are mentioned, most models fail to provide detailed guidance.  

Our findings highlight who gets left behind in current LLM based accessibility support and underscore the need for taxonomy aware training and evaluation practices. By releasing our dataset, evaluation pipeline, and analysis scripts, we aim to catalyze future research toward responsible and inclusive language technologies that better align with global accessibility standards.

\section{Limitations and Future Work}
Our study has several limitations. First, the dataset is relatively small in scale; while sufficient for initial analysis, it will need to be expanded to cover a broader range of domains and disability-related contexts. Second, we restrict our evaluation to English-language responses and general-purpose questions, which may overlook context-specific or culturally grounded accessibility practices. Finally, our Depth@3 scoring, operationalized via LLM-as-Judge, provides consistency but also inherits the limitations of automated evaluators, particularly when nuanced human judgment is required. 

Future work will explore multilingual benchmarks, conversational accessibility scenarios, and human-in-the-loop validation to strengthen reliability. Additionally, integrating factual accuracy and real-world usability into the evaluation framework can provide a more comprehensive assessment of accessibility-aware LLM behavior.

\bibliography{custom}

\appendix

\section{Additional Experimental Details}
\label{sec:appendix}
\subsection{Prompt Design for LLM-as-Judge Evaluation}.
\label{subsec:llm_judge}
Evaluation using a secondary LLM as the judge. The judge processes the question and model-generated response using the following structured instruction:

\begin{figure}[H]
\centering
\begin{tcolorbox}[colback=gray!10, colframe=black, boxrule=0.1pt]
\footnotesize
\scriptsize
Your goal is given a question and its corresponding response from a language model, check whether the response demonstrates inclusive awareness of various types of disabilities, or if it narrowly focuses on only a few.\newline

\textbf{Disability Categories:}
\begin{enumerate}[noitemsep, topsep=0pt]
    \item Vision Impairments
    \item Hearing Impairments
    \item Speech Impairments
    \item Mobility Impairments
    \item Neurological Disorders
    \item Genetic and Developmental Disorders
    \item Learning Disorders
    \item Sensory Processing and Cognitive Disorders
    \item Mental Health and Behavioral Disorders\newline
\end{enumerate}

Step 1. Extract Content per Category\\
For each disability category, extract the specific part of the response that applies to it. If no relevant content is found in the response for a
particular category, clearly state: "Not applicable".\newline

Step 2. Assign Depth score\\
For each category, assign a Depth Score (0–3) based on the quality of detail:

\begin{enumerate}[noitemsep, topsep=0pt]
\item 0 = Not mentioned → no relevant content.
\item 1 = Shallow → only a name-drop or vague mention.\newline
    Example: "People with vision impairments exist."
\item 2 = Moderate → one concrete example or specific detail.\newline
    Example: "Vision impairments may require screen readers."
\item 3 = Deep → multiple details, examples, or nuanced explanation.\newline
    Example: "People with vision impairments may rely on screen readers, braille devices, tactile maps, or high-contrast displays."\newline
\end{enumerate}

Depth measures whether the response provides meaningful detail or just a superficial mention.\newline

\textbf{Output format}\newline
Return JSON in the following format:

\begin{verbatim}
{
  "Vision Impairments": {
    "snippet": "...",
    "depth": 0
  },
  "Hearing Impairments": {
    "snippet": "...",
    "depth": 0
  },
  "Speech Impairments": {
    "snippet": "Not applicable",
    "depth": 0
  },
  "Mobility Impairments": {
    "snippet": "...",
    "depth": 0
  },
  "Neurological Disorders": {
    "snippet": "...",
    "depth": 0
  },
  "Genetic and Developmental Disorders": {
    "snippet": "Not applicable",
    "depth": 0
  },
  "Learning Disorders": {
    "snippet": "...",
    "depth": 0
  },
  "Sensory Processing and Cognitive Disorders": {
    "snippet": "Not applicable",
    "depth": 0
  },
  "Mental Health and Behavioral Disorders": {
    "snippet": "...",
    "depth": 0
  }
}
\end{verbatim}

\textbf{Input Template:}
\begin{verbatim}
Question:
##Question##

Response:
##Response##
\end{verbatim}
\end{tcolorbox}
\caption{Inclusivity evaluation prompt}
\label{fig:inclusivity_evaluation_prompt}
\end{figure}

\subsection{Prompt Design for improved accessibility awareness}.
\label{subsec:prompt_improve_accessibility_awareness}

\begin{figure}[H]
\centering
\begin{tcolorbox}[colback=gray!10, colframe=black, boxrule=0.1pt]
\footnotesize
\scriptsize
\textbf{Your goal:} You are an accessibility expert. Provide an inclusive, comprehensive, and practical response to the given question.\newline

\textbf{Guidelines:}
\begin{enumerate}[noitemsep, topsep=0pt]
    \item Cover all major disability categories where relevant:
    \begin{itemize}[noitemsep, topsep=0pt, leftmargin=1.5em]
        \item Vision Impairments
        \item Hearing Impairments
        \item Speech Impairments
        \item Mobility Impairments
        \item Neurological Disorders
        \item Genetic and Developmental Disorders
        \item Learning Disorders
        \item Sensory Processing and Cognitive Disorders
        \item Mental Health and Behavioral Disorders
    \end{itemize}
    \item Ensure practical, actionable advice or examples for each category when relevant.
    \item Maintain inclusive language and avoid stereotypes or assumptions.
\end{enumerate}

\textbf{Input Template:}
\begin{verbatim}
Question:
##Question##
\end{verbatim}
\end{tcolorbox}
\caption{Accessibility-aware response generation prompt}
\label{fig:accessibility_response_prompt}
\end{figure}

\subsection{Depth Results at 0, 1, 2}.
\label{subsec:depth_other}

\begin{table*}[ht]
\centering
\caption{Depth@0 (fraction of responses with depth=3) by model and disability category.}
\label{tab:depth0}
\resizebox{\textwidth}{!}{%
\begin{tabular}{lccccccccc}
\hline
\textbf{Model} & \textbf{Vision} & \textbf{Hearing} & \textbf{Speech} & \textbf{Mobility} & \textbf{Neuro} & \textbf{Gen/Dev} & \textbf{Learning} & \textbf{Sens/Cog} & \textbf{Mental} \\
\hline
Llama-3.1-8B-Instruct        & 0.229 & 0.343 & 0.714 & 0.229 & 0.629 & 0.800 & 0.486 & 0.543 & 0.600 \\
Llama-3.2-1B-Instruct        & 0.086 & 0.200 & 0.771 & 0.114 & 0.543 & 0.800 & 0.371 & 0.514 & 0.457 \\
Llama-3.2-3B-Instruct        & 0.257 & 0.343 & 0.800 & 0.314 & 0.543 & 0.771 & 0.486 & 0.400 & 0.571 \\
Llama-3.3-70B-Instruct       & 0.200 & 0.257 & 0.800 & 0.200 & 0.400 & 0.600 & 0.400 & 0.371 & 0.514 \\
Ministral-8B-Instruct-2410    & 0.171 & 0.229 & 0.771 & 0.114 & 0.429 & 0.857 & 0.400 & 0.371 & 0.543 \\
Phi-4-mini-instruct           & 0.143 & 0.229 & 0.800 & 0.171 & 0.543 & 0.714 & 0.457 & 0.457 & 0.629 \\
Qwen2.5-0\_5B-Instruct       & 0.229 & 0.286 & 0.829 & 0.229 & 0.657 & 0.800 & 0.629 & 0.600 & 0.629 \\
Qwen2.5-14B-Instruct         & 0.200 & 0.314 & 0.829 & 0.257 & 0.429 & 0.800 & 0.457 & 0.486 & 0.600 \\
Qwen2.5-1\_5B-Instruct       & 0.257 & 0.400 & 0.743 & 0.286 & 0.657 & 0.857 & 0.600 & 0.514 & 0.771 \\
Qwen2.5-32B-Instruct         & 0.114 & 0.200 & 0.657 & 0.171 & 0.514 & 0.857 & 0.457 & 0.457 & 0.714 \\
Qwen2.5-3B-Instruct          & 0.229 & 0.371 & 0.714 & 0.200 & 0.629 & 0.857 & 0.543 & 0.486 & 0.771 \\
Qwen2.5-72B-Instruct         & 0.143 & 0.229 & 0.743 & 0.257 & 0.571 & 0.800 & 0.457 & 0.343 & 0.657 \\
Qwen2.5-7B-Instruct          & 0.143 & 0.343 & 0.771 & 0.171 & 0.514 & 0.571 & 0.457 & 0.400 & 0.629 \\
Cohere command-a-03-2025     & 0.171 & 0.229 & 0.686 & 0.171 & 0.400 & 0.714 & 0.429 & 0.400 & 0.371 \\
Cohere command-r                     & 0.114 & 0.229 & 0.886 & 0.143 & 0.343 & 0.714 & 0.429 & 0.429 & 0.486 \\
GPT-4.1                      & 0.086 & 0.200 & 0.743 & 0.114 & 0.343 & 0.600 & 0.314 & 0.371 & 0.486 \\
Haiku 3.5                   & 0.257 & 0.286 & 0.743 & 0.257 & 0.457 & 0.571 & 0.343 & 0.429 & 0.371 \\ \hline
\end{tabular}%
}
\end{table*}

\begin{table*}[ht]
\centering
\caption{Depth@1 (fraction of responses with depth=3) by model and disability category.}
\label{tab:depth1}
\resizebox{\textwidth}{!}{%
\begin{tabular}{lccccccccc}
\hline
\textbf{Model} & \textbf{Vision} & \textbf{Hearing} & \textbf{Speech} & \textbf{Mobility} & \textbf{Neuro} & \textbf{Gen/Dev} & \textbf{Learning} & \textbf{Sens/Cog} & \textbf{Mental} \\
\hline
Llama-3.1-8B-Instruct & 0.200 & 0.200 & 0.143 & 0.286 & 0.257 & 0.143 & 0.286 & 0.343 & 0.286 \\
Llama-3.2-1B-Instruct & 0.343 & 0.257 & 0.086 & 0.371 & 0.314 & 0.143 & 0.286 & 0.257 & 0.429 \\
Llama-3.2-3B-Instruct & 0.200 & 0.171 & 0.086 & 0.200 & 0.286 & 0.171 & 0.257 & 0.286 & 0.257 \\
Llama-3.3-70B-Instruct & 0.200 & 0.200 & 0.029 & 0.200 & 0.400 & 0.314 & 0.286 & 0.429 & 0.257 \\
Ministral-8B-Instruct-2410 & 0.314 & 0.286 & 0.143 & 0.286 & 0.371 & 0.086 & 0.400 & 0.343 & 0.257 \\
Phi-4-mini-instruct & 0.371 & 0.371 & 0.086 & 0.286 & 0.286 & 0.200 & 0.229 & 0.343 & 0.314 \\
Qwen2.5-0\_5B-Instruct & 0.486 & 0.429 & 0.057 & 0.343 & 0.343 & 0.171 & 0.257 & 0.343 & 0.286 \\
Qwen2.5-14B-Instruct & 0.229 & 0.343 & 0.114 & 0.286 & 0.371 & 0.114 & 0.371 & 0.343 & 0.257 \\
Qwen2.5-1\_5B-Instruct & 0.429 & 0.400 & 0.200 & 0.314 & 0.314 & 0.114 & 0.286 & 0.429 & 0.171 \\
Qwen2.5-32B-Instruct & 0.286 & 0.286 & 0.171 & 0.114 & 0.371 & 0.114 & 0.229 & 0.286 & 0.200 \\
Qwen2.5-3B-Instruct & 0.286 & 0.229 & 0.114 & 0.200 & 0.286 & 0.143 & 0.229 & 0.257 & 0.143 \\
Qwen2.5-72B-Instruct & 0.371 & 0.371 & 0.086 & 0.229 & 0.229 & 0.171 & 0.257 & 0.371 & 0.171 \\
Qwen2.5-7B-Instruct & 0.343 & 0.314 & 0.114 & 0.229 & 0.371 & 0.343 & 0.314 & 0.400 & 0.314 \\
Cohere command-a-03-2025 & 0.171 & 0.171 & 0.114 & 0.200 & 0.314 & 0.229 & 0.229 & 0.286 & 0.371 \\
Cohere command-r & 0.400 & 0.343 & 0.057 & 0.257 & 0.457 & 0.257 & 0.257 & 0.286 & 0.229 \\
GPT-4.1 & 0.429 & 0.371 & 0.114 & 0.286 & 0.286 & 0.286 & 0.314 & 0.314 & 0.314 \\
Haiku 3.5 & 0.371 & 0.343 & 0.143 & 0.286 & 0.314 & 0.400 & 0.400 & 0.371 & 0.514 \\ \hline
\end{tabular}%
}
\end{table*}

\begin{table*}[ht]
\centering
\caption{Depth@2 (fraction of responses with depth=3) by model and disability category.}
\label{tab:depth2}
\resizebox{\textwidth}{!}{%
\begin{tabular}{lccccccccc}
\hline
\textbf{Model} & \textbf{Vision} & \textbf{Hearing} & \textbf{Speech} & \textbf{Mobility} & \textbf{Neuro} & \textbf{Gen/Dev} & \textbf{Learning} & \textbf{Sens/Cog} & \textbf{Mental} \\
\hline
Llama-3.1-8B-Instruct & 0.34 & 0.40 & 0.11 & 0.34 & 0.11 & 0.06 & 0.20 & 0.09 & 0.11 \\
Llama-3.2-1B-Instruct & 0.31 & 0.49 & 0.14 & 0.26 & 0.14 & 0.06 & 0.34 & 0.20 & 0.11 \\
Llama-3.2-3B-Instruct & 0.31 & 0.43 & 0.11 & 0.31 & 0.17 & 0.06 & 0.26 & 0.29 & 0.17 \\
Llama-3.3-70B-Instruct & 0.37 & 0.49 & 0.17 & 0.37 & 0.20 & 0.09 & 0.31 & 0.17 & 0.23 \\
Ministral-8B-Instruct-2410 & 0.31 & 0.34 & 0.09 & 0.40 & 0.20 & 0.06 & 0.20 & 0.29 & 0.20 \\
Phi-4-mini-instruct & 0.29 & 0.34 & 0.09 & 0.34 & 0.17 & 0.09 & 0.31 & 0.17 & 0.06 \\
Qwen2.5-0\_5B-Instruct & 0.23 & 0.29 & 0.11 & 0.37 & 0.00 & 0.03 & 0.11 & 0.06 & 0.09 \\
Qwen2.5-14B-Instruct & 0.34 & 0.31 & 0.06 & 0.26 & 0.20 & 0.09 & 0.14 & 0.17 & 0.14 \\
Qwen2.5-1\_5B-Instruct & 0.14 & 0.20 & 0.06 & 0.40 & 0.03 & 0.03 & 0.11 & 0.06 & 0.06 \\
Qwen2.5-32B-Instruct & 0.37 & 0.40 & 0.17 & 0.49 & 0.11 & 0.03 & 0.26 & 0.23 & 0.09 \\
Qwen2.5-3B-Instruct & 0.29 & 0.31 & 0.17 & 0.49 & 0.09 & 0.00 & 0.23 & 0.26 & 0.09 \\
Qwen2.5-72B-Instruct & 0.23 & 0.31 & 0.17 & 0.31 & 0.20 & 0.03 & 0.29 & 0.29 & 0.17 \\
Qwen2.5-7B-Instruct & 0.31 & 0.23 & 0.11 & 0.40 & 0.11 & 0.09 & 0.23 & 0.20 & 0.06 \\
Cohere command-a-03-2025 & 0.34 & 0.51 & 0.17 & 0.40 & 0.23 & 0.06 & 0.34 & 0.31 & 0.26 \\
Cohere command-r & 0.20 & 0.34 & 0.03 & 0.34 & 0.20 & 0.03 & 0.29 & 0.29 & 0.26 \\
GPT-4.1 & 0.14 & 0.34 & 0.11 & 0.31 & 0.34 & 0.11 & 0.31 & 0.29 & 0.20 \\
gpt-oss-20b & 0.26 & 0.31 & 0.00 & 0.40 & 0.14 & 0.03 & 0.23 & 0.11 & 0.06 \\
Haiku 3.5 & 0.14 & 0.26 & 0.09 & 0.23 & 0.23 & 0.03 & 0.23 & 0.20 & 0.11 \\ \hline
\end{tabular}%
}
\end{table*}


\end{document}